\documentclass[conference]{IEEEtran}
\IEEEoverridecommandlockouts
\usepackage{cite}
\usepackage{amsmath,amssymb,amsfonts}
\usepackage{graphicx}
\usepackage{stfloats} 
\usepackage{textcomp}
\usepackage{xcolor}
\usepackage{braket}
\usepackage{algorithm}
\usepackage{algpseudocode}
\usepackage{booktabs}
\usepackage{quantikz}
\usepackage{float}
\usepackage{hyperref}
\usepackage[utf8]{inputenc}
\usepackage[T1]{fontenc}
\usepackage{caption}
\usepackage[normalem]{ulem}

\def\BibTeX{{\rm B\kern-.05em{\sc i\kern-.025em b}\kern-.08em
    T\kern-.1667em\lower.7ex\hbox{E}\kern-.125emX}}
\begin{document}

\title{\huge A deep dive into the interplay of structured quantum peaked circuits and infinite temperature correlation functions\\
}

\author{
\begin{minipage}[t]{0.32\textwidth}
    \centering
    Myeongsu Kim \\
    \textit{Department of Computer Science} \\
    \textit{Purdue University} \\
    West Lafayette, USA \\
    kim2167@purdue.edu
\end{minipage}
\hfill
\begin{minipage}[t]{0.32\textwidth}
    \centering
    Manas Sajjan \\
    \textit{Department of Electrical and Computer Engineering} \\
    \textit{North Carolina State University} \\
    Raleigh, USA \\
    msajjan@ncsu.edu
\end{minipage}
\hfill
\begin{minipage}[t]{0.32\textwidth}
    \centering
    Sabre Kais \\
    \textit{Department of Electrical and Computer Engineering} \\
    \textit{North Carolina State University} \\
    Raleigh, USA \\
    skais@ncsu.edu
\end{minipage}
}

\maketitle

\begin{abstract}
Random quantum circuits have been extensively explored for quantum supremacy demonstrations \cite{rc1, rc2, aaronson2023certified, movassagh2020quantum}. However, verifying their output distributions remains challenging \cite{rc2, rc4, rc5}. Here, we propose the infinite-temperature correlation function (ITCF) as a physically meaningful observable for noisy intermediate-scale quantum (NISQ) devices one that can be extracted using engineered circuits rather than relying on fully random constructions. This is realized by leveraging peaked quantum states whose probability distributions are sharply peaked at specific outcomes due to constructive interference thus offering more efficient verifiability and stronger signal observability. Rather than using Haar-random states, which often yield vanishing signals through destructive interference, we construct purposefully biased quantum states using either Grover-based amplitude amplification or shallow structured circuits. These engineered states amplify contributions from relevant operator subspaces, enabling robust detection of non-zero ITCF values that would otherwise be suppressed under random-state sampling. Our results highlight a problem-specific state preparation framework that mitigates signal loss from random averaging and facilitates the detection of physically meaningful observables in NISQ devices. We also discuss future extensions to multi-qubit observables, scrambling diagnostics, and variational circuit optimization, underscoring the broader potential of Peaked States for quantum simulation and verification.
\end{abstract}


\section{Introduction}
Random quantum circuits have played a central role in recent efforts to demonstrate quantum advantage, particularly in the context of sampling tasks believed to be beyond the reach of classical computers~\cite{rc1, aaronson2023certified, movassagh2020quantum}. By drawing from the vast space of random unitary operations, these circuits generate highly entangled states that are exceedingly difficult for classical hardware to simulate. However, verifying these outputs---essential for both complexity-theoretic claims and practical certification---remains a formidable hurdle, especially as the system size grows, due to the inherent randomness and lack of physically interpretable structure in the generated states~\cite{rc2, rc4, rc5}. Moreover, recent studies show that output distributions undergo sharp phase transitions with depth and noise, further reinforcing their classical inaccessibility~\cite{morvan2024phase}.

This verification challenge raises a broader question: how can we extract physically meaningful observables from near-term, noisy intermediate-scale quantum (NISQ) devices, when random sampling often yields exponentially small signals? A promising direction is the use of \emph{peaked circuits}—quantum circuits that bias probability distributions toward specific bitstrings. Initially introduced to improve the verifiability of random circuit sampling~\cite{peaked1}, they boost the likelihood of “heavy” outputs in a complexity-theoretic sense. More broadly, this biasing principle motivates new applications: designing circuits that amplify physically informative components of quantum observables in structured and experimentally feasible ways.

One such observable is the \textit{infinite-temperature correlation function} (ITCF), which captures the dynamical relationship between quantum operators at different times under thermal equilibrium at infinite temperature. ITCFs have been widely employed in studies of quantum chaos, thermalization, and information scrambling~\cite{related1, related2}, offering a conceptually clear and hardware-compatible tool for probing many-body quantum dynamics. In this work, we specifically focus on the equal-time case ($t=0$), as it presents an experimentally accessible scenario on current NISQ hardware while still capturing essential aspects of operator correlations. However, estimating ITCFs using Haar-random or uniformly sampled pure states often leads to destructive interference across the Hilbert space, causing the expectation value to vanish exponentially with system size—even under ideal, noiseless conditions. This makes the full-space trace definition of ITCF, while theoretically well-defined, practically inaccessible in realistic experiments where only a limited number of samples can be taken.

To overcome this challenge, we propose a practical estimator that selectively amplifies operator-relevant contributions, thereby avoiding the need to average over the full Hilbert space. This is achieved using peaked quantum states—engineered to concentrate probability on physically informative basis components while preserving the core structure of the ITCF. We explore two complementary strategies: (1) Grover-based amplitude amplification, and (2) shallow, structured circuits tailored to emphasize operator-relevant basis states. While our estimator does not reproduce the exact, unbiased trace over the full Hilbert space, it provides a biased but operationally useful alternative that captures operator-relevant dynamics under realistic constraints. This tradeoff—between theoretical generality and experimental feasibility—defines the scope of our contribution. In contrast to prior work relying on deep or variationally optimized circuits~\cite{peaked1}, our method offers a lightweight and interpretable approach. Experiments on IBM Q devices confirm that structured peaked circuits can robustly amplify relevant correlation signals, offering a practical pathway to probe many-body dynamics in today’s noisy quantum environments.

% In the next sections, we describe the construction of such circuits, analyze their behavior theoretically and experimentally, and discuss extensions to multi-qubit observables, scrambling diagnostics (e.g., out-of-time-order correlators), and hybrid variational protocols that could further enhance peaked-state design.

\section{Background and Problem Setup}
\label{sec:background}

\subsection{Infinite-Temperature Correlation Function}
\label{subsec:itcf}

The infinite-temperature correlation function (ITCF) is a physically meaningful observable that captures correlations between two quantum operators, \( A(t) \) and \( B(0) \). Here, \( A(t) = e^{iHt} A(0) e^{-iHt} \) evolves under the Hamiltonian \( H \), while \( B(0) \) remains static. The ITCF is generally defined as:
\begin{equation}
C_{AB}(t) = \mathrm{Tr}(A(t) B(0) \rho),
\end{equation}
where \( \rho \) is the initial state of the system. In the infinite-temperature limit \( \beta \to 0 \), the state becomes maximally mixed, \( \rho = I / 2^n \), where \( n \) is the number of qubits. Substituting this gives:
\begin{equation}
C_{AB}(t) = \frac{1}{2^n} \mathrm{Tr}(A(t) B(0)).
\end{equation}

In this work, we focus on the equal-time case \( t = 0 \), with further details discussed later.  
Throughout this work, the operators \( A \) and \( B \) were implemented as single-qubit Pauli-\( Z \) observables.

\subsection{Reformulation of the Estimator}
\label{subsec:estimator}

To estimate the ITCF numerically or experimentally, a standard approach is to sample multiple random pure states \( |r_i\rangle \) and average their expectation values \cite{estim_trace}:
\begin{equation}
\tilde{C}_{AB}(t) \approx \frac{1}{2^n} \sum_{i=1}^M \langle r_i | A(t) B(0) | r_i \rangle.
\end{equation}

To reduce signal cancellation, one can define \( B(0) \) as a projector-based observable:
\begin{equation}
B(0) = 2P_\uparrow - I, \quad \text{where } P_\uparrow = |0\rangle\langle0|.
\end{equation}
Here, \( P_\uparrow \) denotes a single-qubit projector onto the \( +1 \) eigenstate of the Pauli-\( Z \) operator, i.e., \( |0\rangle\langle 0| \), acting on a qubit that is distinct from those on which \( A(t) \) acts. This separation ensures that the projector selects a meaningful subspace without trivially commuting with the observable. It satisfies \( P_\uparrow^2 = P_\uparrow \), and leads to the following reformulated estimator:
\begin{equation}
\tilde{C}_{AB}(t) \approx \frac{1}{2^n} \sum_{i=1}^M \left( 2 \langle r_i | P_\uparrow A(t) P_\uparrow | r_i \rangle - \langle r_i | A(t) | r_i \rangle \right).
\end{equation}

According to Ref.~\cite{canonical}, random states drawn from unitary-invariant ensembles approximate the infinite-temperature mixed state and satisfy:
\begin{align}
\mathbb{E}_r[\langle r | A(t) | r \rangle] &= \frac{1}{2^n} \mathrm{Tr}(A(t)), \\
\mathrm{Var}_r[\langle r | A(t) | r \rangle] &\sim \mathcal{O}\left( \frac{1}{2^n} \right).
\end{align}

Hence, convergence improves with system size, and in some cases—particularly when variance is sufficiently suppressed due to concentration of measure—a single sample may suffice. This behavior is similar to that of thermal pure quantum (TPQ) states~\cite{canonical}, though our peaked states are not formally TPQ, but are engineered to amplify operator contributions in ITCF estimation. Therefore, when the system is large enough, the estimator simplifies to:
\begin{equation}
\tilde{C}_{AB}(t) \approx \frac{1}{2^{n-1}} \langle r | P_\uparrow A(t) P_\uparrow | r \rangle - \frac{1}{2^n} \langle r | A(t) | r \rangle.
\end{equation}

Both terms involve time evolution and can be expanded as:
\begin{align}
\langle r | P_\uparrow A(t) P_\uparrow | r \rangle &= \langle r | P_\uparrow V(-t) A(0) V(t) P_\uparrow | r \rangle, \\
\langle r | A(t) | r \rangle &= \langle r | V(-t) A(0) V(t) | r \rangle.
\end{align}
Here, \( V(t) = e^{-iHt} \) is the time-evolution operator. For the first term, inserting resolutions of the identity in the computational basis \( \{ |z_{j,k}\rangle \} \) gives:
\begin{equation}
\sum_{j,k} \langle r | P_\uparrow V(-t) | z_j \rangle \langle z_j | A(0) | z_k \rangle \langle z_k | V(t) P_\uparrow | r \rangle.
\end{equation}

Assuming that \( A(0) \) is diagonal in the computational basis, the off-diagonal terms \( j \ne k \) vanish, and Equation (11) can be simplified to:

\begin{equation}
\sum_j \langle r | P_\uparrow V(-t) | z_j \rangle \langle z_j | A(0) | z_j \rangle \langle z_j | V(t) P_\uparrow | r \rangle.
\end{equation}

Furthermore, for \( t \to 0 \), the time-evolution operator can be approximated as:
\begin{equation}
V(t) \approx I - iHt + \mathcal{O}(t^2).
\end{equation}

By retaining terms up to the zeroth order in the time in the above expansion, we obtain the simplified form:

\begin{equation}
\langle r | P_\uparrow A(t) P_\uparrow | r \rangle \approx \sum_j |\langle r | P_\uparrow | z_j \rangle|^2 \langle z_j | A(0) | z_j \rangle.
\end{equation}

By the same logic, Equation (10) can also be manipulated given the structure \( A(t) = V(-t) A(0) V(t) \). If we again expand terms in $V(t)$ up to zeroth order in time then the quantity transforms to \( \langle r | A(0) | r \rangle \). We leave it as that without expanding it further in the computational basis. Consequently, the full estimator can be approximated as:
\begin{equation}
\tilde{C}_{AB}(0) \approx \frac{1}{2^{n-1}} \sum_j |\langle r | P_\uparrow | z_j \rangle|^2 \langle z_j | A(0) | z_j \rangle - \frac{1}{2^n} \langle r | A(0) | r \rangle.
\end{equation}

\subsection{Limitations of Haar Sampling}
\label{subsec:haar-limitations}

Although Haar sampling is theoretically justified, it faces practical limitations on real quantum hardware:

\begin{itemize}

  \item \textbf{Cancellation of contribution}: If \( \mathrm{Tr}(AB) = 0 \), then samples with positive and negative contributions cancel out on average. Even when a non-zero ITCF value is desired, this condition does not allow us to engineer the sampling process to achieve constructive interference in a physically meaningful way.
  \item \textbf{Large Hilbert space:} As the system size increases, individual contributions scale as \( \sim 1/D \), leading to a poor signal-to-noise ratio (SNR).
  \item \textbf{Limited shots on NISQ devices:} In practice, measurement repetitions are constrained, and statistical convergence becomes difficult.
\end{itemize}

Thus, although Haar sampling yields unbiased estimators in theory, it is often inefficient in realistic experimental settings.

\begin{figure*}[ht]
\centering
\includegraphics[width=0.65\textwidth]{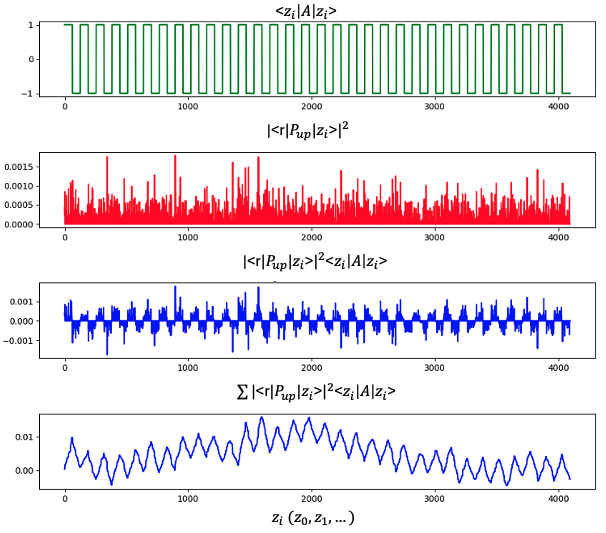}
\caption{
Haar-random state sampling result (classical simulation). From top to bottom, each row shows:
(1) the diagonal elements of the observable \( \langle z | A(0) | z \rangle \);
(2) the probability weights \( |\langle r | P_\uparrow | z \rangle|^2 \);
(3) the signed product \( |\langle r | P_\uparrow | z \rangle|^2 \langle z | A(0) | z \rangle \);
(4) the cumulative sum \( \sum_j |\langle r | P_\uparrow | z_j \rangle|^2 \langle z_j | A(0) | z_j \rangle \).
The weak signal arises due to cancellation between positive and negative contributions. \textbf{The operators $A$ and $B$ were both implemented as Pauli-$Z$ observables in all experiments.}
}
\label{fig:classic_uniform}
\end{figure*}

\subsection{Peaked States}
\label{subsec:peaked}

To overcome these limitations, we propose \textit{peaked quantum states}—quantum states designed to concentrate probability on computational basis states that contribute significantly to the observable. The estimator:
\begin{equation}
\tilde{C}_{AB}(0) \approx \sum_j |\langle r | P_\uparrow | z_j \rangle|^2 \langle z_j | A(0) | z_j \rangle
\end{equation}
is a weighted average over computational basis states \( |z_j\rangle \), where the weights are given by \( |\langle r | P_\uparrow | z_j \rangle|^2 \). If \( \langle z_j | A(0) | z_j \rangle \) includes both positive and negative values, destructive interference may reduce the signal. Thus, peaked states should be designed to concentrate probability on certain \( z_j \) that contribute constructively to the expectation value. The second term with \( \langle r | A(0) | r \rangle \) is omitted, as its contribution can be effectively controlled through the design of the peaked state, allowing us to focus on the dominant projected component in the estimation.

\paragraph{Definition.} A peaked state is a quantum state in which probability is concentrated on computational basis states that significantly influence the observable.

\begin{figure*}[t]
\centering
\includegraphics[width=0.65\linewidth]{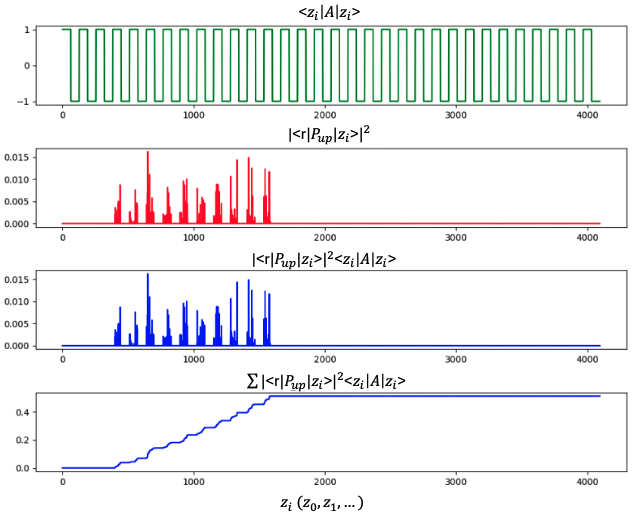}
\caption{
Peaked-state sampling result (classical simulation). Here, probability is concentrated on constructive regions, leading to clear signal amplification.
}
\label{fig:classic_uniform}
\end{figure*}

\paragraph{Quantitative metrics}
To characterize peakedness, we introduce two metrics:
\begin{itemize}
  \item \textbf{Support-overlap score}:
  \[
  S_A(|\Psi\rangle) = \sum_{z: a_z \ne 0} |\langle z | \Psi \rangle|^2
  \]
  \item \textbf{Biased expectation ratio}:
  \[
  E_A(|\Psi\rangle) = \frac{\sum_z a_z p(z)}{\sum_z |a_z| p(z)},
  \]
  \[
  \quad
  a_z = \langle z | A | z \rangle, \quad p(z) = |\langle z | \Psi \rangle|^2
  \]
\end{itemize}

While Haar-random states typically lead to vanishing expectation values due to destructive interference, peaked states concentrate probability on specific computational basis elements that contribute constructively. As a result, the estimator can produce significantly larger magnitude values (\(\left[ \min\left( \langle z_j | A(0) | z_j \rangle \right), \max\left( \langle z_j | A(0) | z_j \rangle \right) \right]\)), enhancing signal strength without requiring extensive averaging.

\section{Circuit Design Strategy}
\label{sec:circuits}

\subsection{Objectives}

As discussed in the previous section, Haar-random states yield statistically valid estimates of the ITCF but are often inefficient on NISQ devices due to limited measurement shots and hardware noise. Moreover, no systematic protocol currently exists for constructing quantum states from random precursors that can consistently yield user-tailored, non-zero correlations. To address this challenge, we propose a systematic protocol for constructing quantum circuits that generate peaked states—quantum states in which probability is concentrated on specific computational basis elements that contribute strongly to the observable. This is achieved by biasing the probability amplitudes toward basis states \( z_j \) for which \( \langle z_j | A(0) | z_j \rangle \) has large magnitude or aligned sign.

We consider two complementary circuit design strategies: one based on Grover-style amplitude amplification, and the other using a shallow, NISQ-friendly architecture.
In practice, real NISQ hardware often allows only a small number of shots (e.g., a few thousand) and suffers from gate errors on multi-qubit operations, making it crucial to keep circuit depth manageable \cite{waring2024noise}.

\subsection{Grover-Based Circuit: Structured Peaking}
\label{subsec:grover}

Grover's algorithm provides a well-known framework for amplifying the amplitudes of bit-strings satisfying a certain condition.
In our context, we use a Grover-like procedure to amplify computational basis states \(z_j\) that have favorable contributions to the ITCF.
However, the circuit can become relatively deep if we apply multiple iterations, so we must carefully balance the desired amplification gain with noise accumulation on a NISQ device.

\begin{algorithm}[t]
\caption{Grover Peaked Circuit}
\textbf{Input:} \(n\)-qubit system, oracle condition \(f(z_j)\)
(e.g., a threshold or sign check for \(\langle z_j \mid A(0)\mid z_j\rangle\))\\
\textbf{Output:} Peaked quantum state \(\ket{r}\)
\begin{algorithmic}[1]
\State Initialize all qubits to \(\ket{0}^{\otimes n}\)
\State Apply Hadamard gates to all qubits to create uniform superposition
\For{\(k = 1 \) to \(T\)}
    \State Apply oracle \(O_f\) to flip the phase of states satisfying \(f(z_j)\)
    \State Apply Grover diffusion operator \(D\)
\EndFor
\State \Return final state \(\ket{r}\)
\end{algorithmic}
\end{algorithm}

\paragraph{Oracle Definition Example}
Suppose we want to amplify basis states \(z_j\) where 
\(\langle z_j \mid A(0)\mid z_j\rangle > 0\).
Then our oracle \(O_f\) can be implemented to flip the phase when this condition is satisfied, 
i.e., when \(a_{z_j} = \langle z_j\mid A(0)\mid z_j\rangle > 0\).
In our case, we create a Boolean function \(f(z_j) = \Theta\bigl(a_{z_j}\bigr)\) (Heaviside step),
and apply a controlled-\(Z\) gate (or equivalent phase flip) conditioned on \(f(z_j) = 1\).

\begin{figure}[ht]
\centering
\begin{quantikz}
\lstick{$\ket{0}$} & \gate{H} & \gate[wires=3]{O_f} & \gate[wires=3]{D} & \qw \\
\lstick{$\ket{0}$} & \gate{H} &                    &                   & \qw \\
\lstick{$\ket{0}$} & \gate{H} &                    &                   & \qw
\end{quantikz}
\caption{Schematic of a single Grover-like iteration. 
The oracle $O_f$ and the diffusion operator $D$ are shown as conceptual blocks spanning all three qubits. }
\label{fig:grover-circuit}
\vspace{-1em}
\end{figure}
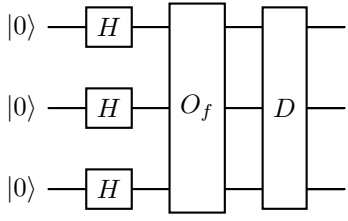

\paragraph{Diffusion Operator Simplification}
The Grover diffusion operator is defined as \( D = 2|\psi\rangle\langle\psi| - I \), where \( |\psi\rangle = H^{\otimes n}|0\rangle \). In standard implementations, this is realized through a combination of Hadamard gates, bit-flips, and multi-controlled phase operations. In this work, we represent it as a conceptual block \( D \), as shown in Figure~\ref{fig:Pj_comparison}, to emphasize generality and abstraction across different system sizes.

\paragraph{NISQ-Specific Example}
On IBM Q devices, for example, the depth of the circuit is typically limited to around 10–15 layers before noise significantly degrades the results \cite{layer_noise1, layer_noise2}. Therefore, the number of Grover iterations \( T \) should be carefully chosen based on system parameters such as the number of qubits, the sparsity of the target state and hardware-specific constraints, including the fidelity and connectivity of the gate. Excessive iterations may result in diminishing returns due to accumulated noise.

% \begin{figure}[h]
% \centering
% \begin{quantikz}
% \lstick{$\ket{0}$} & \gate{H} & \gate[wires=3]{O_f} & \gate[wires=3]{D} & \qw \\
% \lstick{$\ket{0}$} & \gate{H} &                    &                   & \qw \\
% \lstick{$\ket{0}$} & \gate{H} &                    &                   & \qw
% \end{quantikz}
% \caption{Schematic of a single Grover-like iteration. 
% The oracle $O_f$ and the diffusion operator $D$ are shown as conceptual blocks spanning all three qubits. }
% \label{fig:grover-circuit}
% \vspace{-1em}
% \end{figure}

\subsection{Shallow Circuit: Lightweight Peaking}
\label{subsec:shallow}

To create experimentally accessible peaked states with minimal circuit depth, we design a shallow architecture featuring limited entanglement and selective single-qubit rotations.
This approach injects mild bias into the state distribution while avoiding the repeated oracle-diffusion loop of Grover's algorithm.

\begin{algorithm}[h]
\caption{Shallow Peaked Circuit}
\textbf{Input:} \(n\)-qubit system, angle parameters \(\{\theta_i\}\), optional entangling pairs \\
\textbf{Output:} Peaked quantum state \(\ket{r}\)
\begin{algorithmic}[1]
\State Initialize all qubits to \(\ket{0}^{\otimes n}\)
\For{each qubit \(i\)}
    \If{\(i \in \text{HadamardSet}\)}
        \State Apply Hadamard gate \(H\) to qubit \(i\)
    \ElsIf{\(i \in \text{RotationSet}\)}
        \State Apply \(R_y(\theta_i)\) or \(R_z(\phi_i)\) to qubit \(i\)
    \EndIf
\EndFor
\State Apply a small number of CNOT (or CZ) gates between carefully selected qubits
\State (Optional) Apply final single-qubit phase shifts for fine-tuning
\State \Return final state \(\ket{r}\)
\end{algorithmic}
\end{algorithm}

\paragraph{Circuit Structure Example.}
An illustrative design might involve:
\begin{enumerate}
    \item \textbf{Partial Superposition}: Apply Hadamard only to a subset of qubits that significantly affect the observable \(A(0)\), leaving the others in the \(\ket{0}\) state. For example, suppose the observable is \(A(0) = Z_0 \otimes I_1 \otimes Z_2 \otimes I_3 \otimes Z_4\). Then the expectation value \(\langle z_j | A(0) | z_j \rangle\) depends only on qubits 0, 2, and 4. In this case, Hadamard gates are applied only to those three qubits to create superposition, while qubits 1 and 3 are left unentangled in \(\ket{0}\). This reduces circuit depth and avoids introducing noise from qubits that do not influence the measurement.
    \item \textbf{Targeted Bias}: For qubits expected to contribute significantly to the value of \(\langle z_j | A(0) | z_j \rangle\), apply \(R_y(\theta_i)\) rotations to tilt their amplitudes toward \(\ket{0}\) or \(\ket{1}\). Continuing the above example, states with even parity on qubits 0, 2, and 4 (i.e., \(z_0 + z_2 + z_4\) is even) lead to a positive expectation value. To amplify such configurations, \(R_y(\pi/4)\) rotations are applied to qubits 0, 2, and 4. This biases the probability distribution toward favorable basis states and enhances the signal-to-noise ratio in downstream correlation estimates.
    \item \textbf{Minimal Entanglement}: If the observable favors certain correlated patterns (e.g., even parity across selected qubits), a few CNOT gates can be inserted to reinforce this structure. In the example above, applying CNOT(0,2) and CNOT(2,4) ensures that qubits 0, 2, and 4 are likely to take on the same value. This increases the probability weight of favorable \(z_j\) states while maintaining a low overall circuit depth.
\end{enumerate}

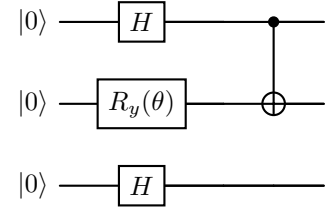
\begin{figure}[H]
\centering
\begin{quantikz}
\lstick{$\ket{0}$} & \gate{H} & \qw & \ctrl{1} & \qw \\
\lstick{$\ket{0}$} & \gate{R_y(\theta)} & \qw & \targ{} & \qw \\
\lstick{$\ket{0}$} & \gate{H} & \qw & \qw & \qw
\end{quantikz}
\caption{Example of a shallow peaked circuit using selective single-qubit rotations and minimal entanglement. Suitable for NISQ devices with gate-depth limitations.}
\label{fig:shallow-circuit}
\vspace{-1em}
\end{figure}

\paragraph{NISQ-Specific Example.}
If hardware connectivity is linear (e.g., qubits 1--2--3--4 in a chain), applying only a few CNOT gates between adjacent qubits can help keep the overall circuit depth within a manageable range (e.g., $\leq 6$). Given that single-qubit gate error rates are typically $\sim 10^{-3}$ and two-qubit gate error rates $\sim 10^{-2}$ \cite{gate_error1, gate_error2}, shallow designs with a small number of entangling operations often retain usable signal fidelity---especially for small to moderate numbers of qubits and when a few thousand measurement shots are available.

\subsection{Qualitative Analysis}

Both strategies aim to control the distribution

\begin{gather}
\langle r \mid P_{\uparrow} A(0) \, P_{\uparrow} \mid r \rangle 
\approx \sum_j p_j \, a_j, \\
p_j = \bigl|\langle r \mid P_{\uparrow} \mid z_j\rangle\bigr|^2,\quad
a_j = \langle z_j \mid A(0)\mid z_j\rangle.
\end{gather}

The design goal is to align \(p_j\) with the sign and magnitude of \(a_j\).
Concretely, if \(a_j > 0\), we want higher \(p_j\); if \(a_j < 0\), we suppress \(p_j\).
\\
\subsubsection*{\textbf{Peakedness Metrics}}\hfill\\

(i) Support-overlap score:
\begin{equation}
S_A(\ket{r}) 
= \sum_{z: a_z \ne 0} \bigl|\langle z\mid P_{up}\mid r \rangle\bigr|^2
\label{eq:support-score}
\end{equation}

(ii) Biased expectation ratio:
\begin{equation}
E_A(\ket{r}) 
= \frac{\sum_z a_z \,p(z) = C_{AB}}{\sum_z |a_z| \, p(z)}.
\label{eq:biased-ratio}
\end{equation}

To further quantify the peakedness of each state preparation strategy, we computed the support-overlap score \( S_A \), the estimated correlation value \( C_{AB} \), and the biased expectation ratio \( E_A = C_{AB}/S_A \).

\begin{table}[H]
\centering
\resizebox{0.8\columnwidth}{!}{%
\begin{tabular}{lccc}
\toprule
\textbf{Method} & \textbf{\(S_A\)} & \textbf{\(C_{AB}\)} & \textbf{\(E_A\)} \\
\midrule
Haar            & 0.4871           & 0.0002              & 0.0005           \\
Grover (T = 3)  & 0.5709           & 0.3920              & 0.6867           \\
Shallow         & 0.9835           & 0.5658              & 0.5753           \\
\bottomrule
\end{tabular}
}
\caption{Comparison of peakedness metrics across state preparation methods.}
\label{tab:peakedness_metrics}
\end{table}

As shown in Table~\ref{tab:peakedness_metrics}, both the Grover and shallow circuits yield significantly higher values across all metrics compared to Haar-random sampling, indicating a much stronger alignment with the relevant operator subspace. While the Grover and shallow methods differ in structure and depth, we note that such engineered circuits can, in principle, be tuned to optimize \( E_A \), and a full exploration of their limits is left for future work.

\subsection{Implementation Trade-Offs}

\begin{table}[H]
\centering
\resizebox{1.0\columnwidth}{!}{%
\begin{tabular}{lcc}
\toprule
\textbf{Aspect} & \textbf{Grover-Based Circuit} & \textbf{Shallow Circuit}\\
\midrule
Circuit depth         & Typically deep (multiple iterations)
                     & Generally low (few gates) \\
Amplification strength & Strong (theoretically optimal)
                     & Moderate/Heuristic \\
NISQ compatibility    & Lower (noise accumulates)
                     & Higher (less complex) \\
Design flexibility    & Less flexible
                     & Highly tunable \\
Randomness    & Medium (oracle + measurement)
                     & Low–Medium \\
\bottomrule
\end{tabular}%
}
\caption{Comparison of Grover and Shallow peaked circuit strategies under NISQ constraints.}
\label{tab:tradeoffs}
\end{table}

\subsection{Complementary Use Strategy and Next Steps}

In summary, these two peaking methods address distinct trade-offs:

\begin{itemize}
\item \textbf{Shallow circuits:} Ideal for quick experimentation and lower-noise runs, requiring minimal resources while offering only modest amplitude amplification.
\item \textbf{Grover-based circuits:} Capable of achieving stronger amplification in theory, but more sensitive to noise and requiring careful oracle construction and iteration tuning.
\end{itemize}

These strategies provide complementary tools for peaked state preparation, allowing practitioners to adapt to the precision and resource constraints of available quantum hardware. In the next section, we demonstrate how each approach performs under realistic shot and gate error conditions, providing guidance on when to favor shallow versus Grover-based strategies for peaked state generation.

\section{Experimental and Simulation Results}
\label{sec:results}

\subsection{Simulation and Experimental Setup}

As shown in Section~\ref{sec:background}, the final circuit construction does not depend on the specific form of the Hamiltonian. Therefore, we assume a 12-qubit system for all experiments, and the observable of interest is the equal-time Infinite Temperature Correlation Function (ITCF). In all experiments, we used Pauli-$Z$ operators for both $A$ and the observable associated with $B$, each applied to a single qubit.

We evaluate the performance of two different state preparation strategies:

\textbf{Grover Circuit.}  
To create peaked states, we implement Grover-based amplitude amplification circuits and simulate them for various iteration counts \( T \in \{1, 2, \dots, 10\} \). For quantum hardware experiments, we focus on \( T = 3 \),  
as it provides a reasonable trade-off between signal amplification and circuit depth:
\textit{T = 3} was chosen as an initial attempt to balance signal amplification and circuit complexity, with further experiments planned under various conditions.

\textbf{Shallow Circuit.}  
As a baseline, we also include a shallow circuit with fixed depth. While it does not implement Grover-style amplitude amplification, it may still induce implicit biases in the probability amplitudes through its structure.

All experiments were conducted on a 12-qubit system with 8192 shots.
\textit{Note: The ITCF values reported in this paper are computed without the normalization factor \( \frac{1}{D} \), using the expression:
\[
\sum_j |\langle r | P_{\text{up}} | z_j \rangle|^2 \langle z_j | A(0) | z_j \rangle
\]  
to focus on relative comparisons between circuit types rather than absolute values.}

\subsection{Classical Simulation Results}

We simulated the Grover circuit for increasing values of \( T \), and observed that the estimated ITCF values improved gradually.  
In particular, ITCF values close to 0.95 were obtained for sufficiently large \( T \), indicating effective signal enhancement through peaked state construction.  
The shallow circuit also consistently outperformed Haar-random sampling in these simulations.

\subsection{Quantum Hardware Results}

\begin{figure*}[t]
\centering
\includegraphics[width=0.65\textwidth]{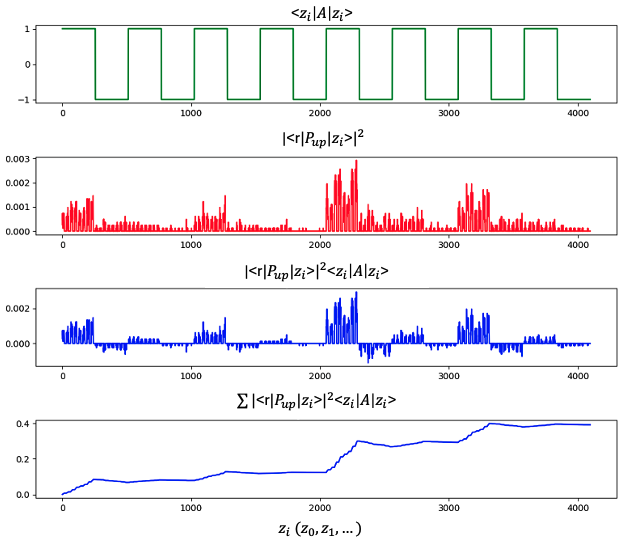}
\caption{ITCF estimates from Grover-based circuit with \( T = 3 \) on IBM Q hardware (\texttt{ibm\_brisbane}, \texttt{ibm\_kyiv}, \texttt{ibm\_sherbrooke}). Consistent performance observed across devices.}
\label{fig:grover-hw}
\end{figure*}

\begin{figure*}[t]
\centering
\includegraphics[width=0.65\textwidth]{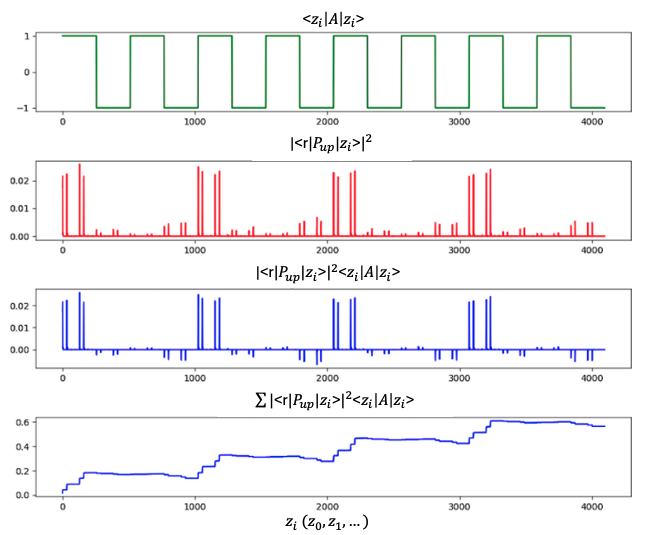}
\caption{ITCF estimates from shallow circuit on IBM Q hardware. Stronger signals observed compared to Grover circuit across all devices.}
\label{fig:shallow-hw}
\end{figure*}

We executed both the Grover circuit (\( T = 3 \)) and the shallow circuit on multiple IBM Q devices, including \texttt{ibm\_brisbane}, \texttt{ibm\_kyiv}, and \texttt{ibm\_sherbrooke}. All experiments were performed with 8192 shots. Grover circuits yielded ITCF signals around 0.38, and these values remained consistent across all hardware platforms, demonstrating the robustness of our method under realistic NISQ noise conditions. The shallow circuit produced even higher ITCF values on the same devices, further highlighting its potential as a hardware-friendly approach.
Figures~\ref{fig:grover-hw} and \ref{fig:shallow-hw} summarize the measured ITCF values.

\subsection{Summary}

From simulation and hardware experiments, we derive the following insights:

\begin{itemize}
    \item \textbf{Grover Circuit:} In simulation, ITCF improves with increasing \( T \), confirming the effectiveness of the amplitude amplification mechanism.  
    On hardware, however, deeper circuits are expected to be more sensitive to noise, which may limit practical benefits.

    \item \textbf{Shallow Circuit:} Despite its simplicity, the shallow circuit achieved strong ITCF signals on real devices, indicating its promise as a noise-resilient and resource-efficient alternative for NISQ-era platforms.
\end{itemize}

These findings validate our circuit design strategies and highlight the potential of peaked state preparation—even with limited quantum resources—for enhancing correlation measurements in near-term quantum devices.

\section{Discussion}

In this section, we discuss the implications of the simulation and hardware results and outline directions for future investigation regarding peaked-state circuit design.

\subsection*{Key Observations}

\begin{figure*}[t]
\centering
\includegraphics[width=0.75\textwidth]{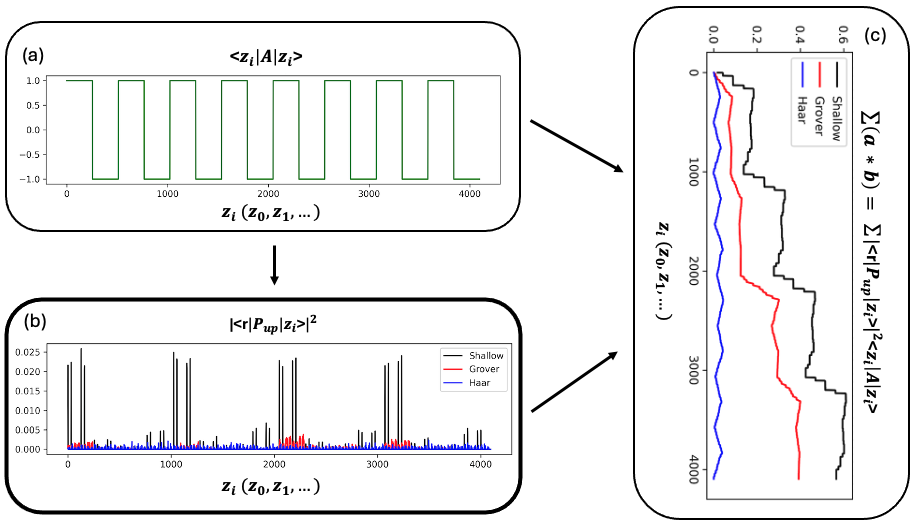}
\caption{Comparison of probability distributions \( |\langle r | P_{\uparrow} | z_j \rangle|^2 \) for three different state preparation strategies. Structured states allocate higher probability to basis states that constructively contribute to the ITCF, while Haar states remain uniformly unstructured.}
% \caption{To be modified.}
\label{fig:Pj_comparison}
\end{figure*}

The Grover-based circuit showed a consistent improvement in ITCF values with increasing iteration count \( T \) in classical simulations. At \( T = 3 \), we also confirmed, through quantum hardware experiments, that the resulting peaked state concentrated probability density on relevant basis states.  
The shallow circuit, despite its fixed and minimal depth, exhibited strong ITCF signals in both classical and hardware experiments.  
These results suggest that both strategies offer robust and efficient alternatives for obtaining meaningful \( C_{AB} \) estimates in NISQ-era quantum devices. Figure~\ref{fig:Pj_comparison} further illustrates the probability distributions \( |\langle r|P_\uparrow|z_j\rangle|^2 \) under each state preparation strategy. While Haar-random states produce a nearly uniform distribution across computational basis states, both the Grover and shallow circuits generate highly structured, peaked profiles that align with operator-relevant subspaces. This confirms that the designed circuits effectively concentrate probability mass on components that contribute to non-zero ITCF values.
\subsection*{Theoretical Significance}

This work demonstrates that peaked-state preparation strategies can serve as practical alternatives to Haar-random sampling, which typically suffers from signal suppression.  
By concentrating probability weight on relevant computational basis states, these circuits enhance correlation function estimation.  
Such methods may contribute to various applications in quantum many-body physics, including interaction diagnostics, information scrambling, and the analysis of quantum chaos.

\subsection*{Applicability and Scalability}

We plan to extend our current approach to larger system sizes and more general classes of observables.  
In particular, the shallow circuit's compatibility with current quantum hardware makes it a promising candidate for large-scale practical implementation on NISQ devices.  
While our current experiments applied \( P_{\text{up}} \) and \( A \) to only a single qubit each, we intend to explore the case where correlation functions of multi-qubit operators are also considered, to further assess the scalability and versatility of the peaked-state approach. Moreover, even beyond condensed matter physical systems where correlation functions between Pauli operators can act as order parameters for distinguishing various many-body phases \cite{sajjan2022magnetic, weidman2024quantum}, even  quantum-machine learning based applications may benefit from the schemes discussed in this manuscript. This is due to the fact that our protocol is entirely general and can be used to engineer states with tailored correlation between certain qubit pairs in the data-encoding stage of usual quantum machine learning workflow \cite{sajjan2022quantum}. There are reports which claim that enhancing the ability to foster quantum correlation within the data-encoding unitary can better exploit the advantages of quantum-enhanced workflows \cite{rath2024quantum,schuld2021effect,kashif2023unified}.
Besides, extension to changing correlation content with external force fields \cite{sajjan2018entangling} or to four-body correlation measurements like out-of-time-ordered correlators \cite{sajjan2023imaginary} can also be undertaken which are important for information scrambling. Thus we see this study lays the foundation for improving correlation measurement efficiency in more complex quantum algorithms, and could support experimental investigations of quantum information scrambling and other non-trivial dynamical behavior.

\subsection*{Limitations and Future Work}

While our study confirms the practical feasibility of peaked-state circuits in NISQ settings, a more systematic theoretical analysis is needed to understand the relationship between circuit structure, output distributions, and correlation enhancement.
Future directions include:
\begin{itemize}
    \item Theoretical analysis of how circuit structure influences probability distributions  
    \item Development of error mitigation and robustness strategies for noisy environments  
    \item Exploration of parameterized and hybrid circuit structures for improved state targeting  
    \item Extension to cases where \( P_{\text{up}} \) and \( A \) are applied to multiple qubits
\end{itemize}

\section{Conclusion}

We presented and evaluated two circuit strategies---Grover-based and shallow designs---for preparing peaked quantum states that improve the estimation of the Infinite Temperature Correlation Function (ITCF). Both methods demonstrated robust performance in classical and hardware experiments, with Grover circuits yielding strong amplification at moderate depths, and shallow circuits offering noise resilience and hardware compatibility.

These results provide a practical alternative to Haar-random sampling for correlation estimation, particularly suited to the constraints of NISQ-era devices. Our work opens the door to scalable approaches for probing many-body dynamics, with implications for quantum information scrambling and quantum advantage benchmarks.

\section*{Acknowledgment}

We acknowledge funding from the Office of Science through the Quantum Science Center (QSC), a National Quantum Information Science Research Center. We also acknowledge discussion and communication with Professor Scott Aaronson on his proposed peaked quantum circuit.

\bibliographystyle{IEEEtran}
\bibliography{IEEE}

\end{document}